\documentclass[twocolumn]{aastex631}

\hypersetup{linkcolor=red,citecolor=blue,filecolor=cyan,urlcolor=magenta}

\usepackage{multirow}
\graphicspath{{./}{figures/}}

\begin{document}

\title{Search for Extended Sources of Neutrino Emission in the Galactic Plane with IceCube}
\affiliation{III. Physikalisches Institut, RWTH Aachen University, D-52056 Aachen, Germany}
\affiliation{Department of Physics, University of Adelaide, Adelaide, 5005, Australia}
\affiliation{Dept. of Physics and Astronomy, University of Alaska Anchorage, 3211 Providence Dr., Anchorage, AK 99508, USA}
\affiliation{Dept. of Physics, University of Texas at Arlington, 502 Yates St., Science Hall Rm 108, Box 19059, Arlington, TX 76019, USA}
\affiliation{CTSPS, Clark-Atlanta University, Atlanta, GA 30314, USA}
\affiliation{School of Physics and Center for Relativistic Astrophysics, Georgia Institute of Technology, Atlanta, GA 30332, USA}
\affiliation{Dept. of Physics, Southern University, Baton Rouge, LA 70813, USA}
\affiliation{Dept. of Physics, University of California, Berkeley, CA 94720, USA}
\affiliation{Lawrence Berkeley National Laboratory, Berkeley, CA 94720, USA}
\affiliation{Institut f{\"u}r Physik, Humboldt-Universit{\"a}t zu Berlin, D-12489 Berlin, Germany}
\affiliation{Fakult{\"a}t f{\"u}r Physik {\&} Astronomie, Ruhr-Universit{\"a}t Bochum, D-44780 Bochum, Germany}
\affiliation{Universit{\'e} Libre de Bruxelles, Science Faculty CP230, B-1050 Brussels, Belgium}
\affiliation{Vrije Universiteit Brussel (VUB), Dienst ELEM, B-1050 Brussels, Belgium}
\affiliation{Department of Physics and Laboratory for Particle Physics and Cosmology, Harvard University, Cambridge, MA 02138, USA}
\affiliation{Dept. of Physics, Massachusetts Institute of Technology, Cambridge, MA 02139, USA}
\affiliation{Dept. of Physics and The International Center for Hadron Astrophysics, Chiba University, Chiba 263-8522, Japan}
\affiliation{Department of Physics, Loyola University Chicago, Chicago, IL 60660, USA}
\affiliation{Dept. of Physics and Astronomy, University of Canterbury, Private Bag 4800, Christchurch, New Zealand}
\affiliation{Dept. of Physics, University of Maryland, College Park, MD 20742, USA}
\affiliation{Dept. of Astronomy, Ohio State University, Columbus, OH 43210, USA}
\affiliation{Dept. of Physics and Center for Cosmology and Astro-Particle Physics, Ohio State University, Columbus, OH 43210, USA}
\affiliation{Niels Bohr Institute, University of Copenhagen, DK-2100 Copenhagen, Denmark}
\affiliation{Dept. of Physics, TU Dortmund University, D-44221 Dortmund, Germany}
\affiliation{Dept. of Physics and Astronomy, Michigan State University, East Lansing, MI 48824, USA}
\affiliation{Dept. of Physics, University of Alberta, Edmonton, Alberta, Canada T6G 2E1}
\affiliation{Erlangen Centre for Astroparticle Physics, Friedrich-Alexander-Universit{\"a}t Erlangen-N{\"u}rnberg, D-91058 Erlangen, Germany}
\affiliation{Physik-department, Technische Universit{\"a}t M{\"u}nchen, D-85748 Garching, Germany}
\affiliation{D{\'e}partement de physique nucl{\'e}aire et corpusculaire, Universit{\'e} de Gen{\`e}ve, CH-1211 Gen{\`e}ve, Switzerland}
\affiliation{Dept. of Physics and Astronomy, University of Gent, B-9000 Gent, Belgium}
\affiliation{Dept. of Physics and Astronomy, University of California, Irvine, CA 92697, USA}
\affiliation{Karlsruhe Institute of Technology, Institute for Astroparticle Physics, D-76021 Karlsruhe, Germany }
\affiliation{Karlsruhe Institute of Technology, Institute of Experimental Particle Physics, D-76021 Karlsruhe, Germany }
\affiliation{Dept. of Physics, Engineering Physics, and Astronomy, Queen's University, Kingston, ON K7L 3N6, Canada}
\affiliation{Department of Physics {\&} Astronomy, University of Nevada, Las Vegas, NV, 89154, USA}
\affiliation{Nevada Center for Astrophysics, University of Nevada, Las Vegas, NV 89154, USA}
\affiliation{Dept. of Physics and Astronomy, University of Kansas, Lawrence, KS 66045, USA}
\affiliation{Centre for Cosmology, Particle Physics and Phenomenology - CP3, Universit{\'e} catholique de Louvain, Louvain-la-Neuve, Belgium}
\affiliation{Department of Physics, Mercer University, Macon, GA 31207-0001, USA}
\affiliation{Dept. of Astronomy, University of Wisconsin{\textendash}Madison, Madison, WI 53706, USA}
\affiliation{Dept. of Physics and Wisconsin IceCube Particle Astrophysics Center, University of Wisconsin{\textendash}Madison, Madison, WI 53706, USA}
\affiliation{Institute of Physics, University of Mainz, Staudinger Weg 7, D-55099 Mainz, Germany}
\affiliation{Department of Physics, Marquette University, Milwaukee, WI, 53201, USA}
\affiliation{Institut f{\"u}r Kernphysik, Westf{\"a}lische Wilhelms-Universit{\"a}t M{\"u}nster, D-48149 M{\"u}nster, Germany}
\affiliation{Bartol Research Institute and Dept. of Physics and Astronomy, University of Delaware, Newark, DE 19716, USA}
\affiliation{Dept. of Physics, Yale University, New Haven, CT 06520, USA}
\affiliation{Columbia Astrophysics and Nevis Laboratories, Columbia University, New York, NY 10027, USA}
\affiliation{Dept. of Physics, University of Oxford, Parks Road, Oxford OX1 3PU, United Kingdom}
\affiliation{Dipartimento di Fisica e Astronomia Galileo Galilei, Universit{\`a} Degli Studi di Padova, 35122 Padova PD, Italy}
\affiliation{Dept. of Physics, Drexel University, 3141 Chestnut Street, Philadelphia, PA 19104, USA}
\affiliation{Physics Department, South Dakota School of Mines and Technology, Rapid City, SD 57701, USA}
\affiliation{Dept. of Physics, University of Wisconsin, River Falls, WI 54022, USA}
\affiliation{Dept. of Physics and Astronomy, University of Rochester, Rochester, NY 14627, USA}
\affiliation{Department of Physics and Astronomy, University of Utah, Salt Lake City, UT 84112, USA}
\affiliation{Oskar Klein Centre and Dept. of Physics, Stockholm University, SE-10691 Stockholm, Sweden}
\affiliation{Dept. of Physics and Astronomy, Stony Brook University, Stony Brook, NY 11794-3800, USA}
\affiliation{Dept. of Physics, Sungkyunkwan University, Suwon 16419, Korea}
\affiliation{Institute of Physics, Academia Sinica, Taipei, 11529, Taiwan}
\affiliation{Dept. of Physics and Astronomy, University of Alabama, Tuscaloosa, AL 35487, USA}
\affiliation{Dept. of Astronomy and Astrophysics, Pennsylvania State University, University Park, PA 16802, USA}
\affiliation{Dept. of Physics, Pennsylvania State University, University Park, PA 16802, USA}
\affiliation{Dept. of Physics and Astronomy, Uppsala University, Box 516, S-75120 Uppsala, Sweden}
\affiliation{Dept. of Physics, University of Wuppertal, D-42119 Wuppertal, Germany}
\affiliation{Deutsches Elektronen-Synchrotron DESY, Platanenallee 6, 15738 Zeuthen, Germany }

\author[0000-0001-6141-4205]{R. Abbasi}
\affiliation{Department of Physics, Loyola University Chicago, Chicago, IL 60660, USA}

\author[0000-0001-8952-588X]{M. Ackermann}
\affiliation{Deutsches Elektronen-Synchrotron DESY, Platanenallee 6, 15738 Zeuthen, Germany }

\author{J. Adams}
\affiliation{Dept. of Physics and Astronomy, University of Canterbury, Private Bag 4800, Christchurch, New Zealand}

\author[0000-0002-9714-8866]{S. K. Agarwalla}
\altaffiliation{also at Institute of Physics, Sachivalaya Marg, Sainik School Post, Bhubaneswar 751005, India}
\affiliation{Dept. of Physics and Wisconsin IceCube Particle Astrophysics Center, University of Wisconsin{\textendash}Madison, Madison, WI 53706, USA}

\author[0000-0003-2252-9514]{J. A. Aguilar}
\affiliation{Universit{\'e} Libre de Bruxelles, Science Faculty CP230, B-1050 Brussels, Belgium}

\author[0000-0003-0709-5631]{M. Ahlers}
\affiliation{Niels Bohr Institute, University of Copenhagen, DK-2100 Copenhagen, Denmark}

\author[0000-0002-9534-9189]{J.M. Alameddine}
\affiliation{Dept. of Physics, TU Dortmund University, D-44221 Dortmund, Germany}

\author{N. M. Amin}
\affiliation{Bartol Research Institute and Dept. of Physics and Astronomy, University of Delaware, Newark, DE 19716, USA}

\author{K. Andeen}
\affiliation{Department of Physics, Marquette University, Milwaukee, WI, 53201, USA}

\author[0000-0003-2039-4724]{G. Anton}
\affiliation{Erlangen Centre for Astroparticle Physics, Friedrich-Alexander-Universit{\"a}t Erlangen-N{\"u}rnberg, D-91058 Erlangen, Germany}

% ...
\collaboration{410}{IceCube Collaboration}

\begin{abstract}

The Galactic plane, harboring a diffuse neutrino flux, is a particularly interesting target to study potential cosmic-ray acceleration sites. Recent gamma-ray
observations by HAWC and LHAASO have presented evidence for multiple Galactic sources
that exhibit a spatially extended morphology and have energy spectra continuing beyond 100
TeV. A fraction of such emission could be produced by interactions of accelerated hadronic cosmic rays, resulting in an excess of high-energy neutrinos clustered near these regions. Using 10
years of IceCube data comprising track-like events that originate from charged-current muon neutrino interactions, we perform a dedicated search for extended
neutrino sources in the Galaxy. We find no evidence for time-integrated neutrino emission from the potential extended sources
studied in the Galactic plane. The most significant location, at 2.6$\sigma$ post-trials, is a 1.7$^\circ$ sized region
coincident with the unidentified TeV gamma-ray source 3HWC J1951+266. We provide strong constraints on
hadronic emission from several regions in the Galaxy.     

\end{abstract}

\section{Introduction} \label{sec:intro}
The search for the sources of cosmic rays is a key area of research in multimessenger astronomy. Cosmic rays up to PeV energies are thought to originate in acceleration sites within the Milky Way known as PeVatrons \citep{2022EPJST.231...27B,Blasi_2013,2019IJMPD..2830022G}. The accelerated cosmic-ray protons interact with the surrounding matter to produce pions, which decay into neutrinos and gamma rays. However, gamma rays may also be produced by leptonic cosmic rays via inverse Compton scattering and/or bremsstrahlung processes. The detection of neutrinos from a Galactic source would provide the definitive evidence for hadronic acceleration therein. The IceCube Neutrino Observatory has been observing a diffuse flux of TeV--PeV energy neutrinos of largely unknown origin \cite{ic10yrs}. While the two most promising candidate sources of astrophysical neutrinos to date are extragalactic \citep{txs,txs2,IceCube:2022der} the near-isotropic flux may contain a small Galactic component as well \citep{IceCube:2017trr,Denton:2017csz}. A more recent IceCube analysis has also reported diffuse neutrino emission from the Galactic plane, compatible with the expectation from cosmic-ray interactions with the interstellar medium \citep{2023icgpdiffuse}.  

The Galactic plane has been extensively surveyed in gamma rays at multi-TeV energies \citep{HESS:2018pbp,2021ApJ...917....6A,2020A&A...642A.190M,2016ApJ...821..129A,2010int..workE...5W}. The High Altitude Water Cherenkov (HAWC) Observatory, and the Large High Altitude Air Shower Observatory (LHAASO) have detected several sources that emit photons at more than 100 TeV \citep{bib:3hwc_catalog,bib:astronomy_HAWC_GalacticPlane56TeV,bib:lhaaso_12}.  At such high energies -- in the so-called Klein-Nishina regime -- gamma-ray emission via inverse Compton scattering is increasingly suppressed \citep{Klein1929}, which means the aforementioned $> 100$ TeV emission could be a signature of hadronic interactions.  Some of these sources exhibit a spatial extension up to $\sim 2$ degrees.  A number of the aforementioned sources have been found in close proximity of high spin-down luminosity pulsars, hinting towards a leptonic origin of the gamma-ray emission \citep{HAWC:2021dtl,Sudoh:2021avj,PhysRevD.105.103013}.  In many cases, the gamma-ray data is not enough to distinguish between hadronic acceleration followed by pion decay, and leptonic interactions at the source \citep{Sudoh:2022sdk,Sudoh:2023qrz}. That is why a comprehensive search for neutrino emission in the galaxy is required. Moreover, diffuse gamma rays with energies between 100 TeV and 1 PeV have also been reported by the Tibet air-shower array \citep{tibetasgcollaborationFirstDetectionSubPeV2021} and LHAASO \citep{LHAASO:2023gne}, further hinting towards the presence of undetected sources that may be accompanied by neutrinos. 

Previous works using IceCube data have analysed several supernova remnants, pulsar wind nebulae and unidentified objects detected in TeV gamma rays as point sources \citep{IceCube:2019lzm,ic10yrs,IceCube:2020svz}, and constrained neutrino emission from the 12 ultra-high-energy sources observed by LHAASO. \citep{IceCube:2022heu}. In this work, we adopt a more extensive, model-independent approach to search for extended sources of neutrino emission in the galactic plane. The last search for extended sources with IceCube only used 1 year of complete detector configuration \citep{IceCube:2014vjc}. This work is an improvement on previous IceCube searches in the Galactic plane in several ways. First, we test for the presence of neutrino sources of multiple possible angular sizes using 9 years of data. Second, we select a catalog of special extended regions of interest (ROI) in the Galactic plane that emit $>$ 50 TeV gamma rays, and test for neutrino emission. We also account for any possible contamination from diffuse emission in the Galactic plane. The paper is structured as follows. Section \ref{sec:ic} briefly reviews the detector and data sample used in the search. Section \ref{sec:ana} describes the analysis details and provides the results of the various searches conducted. Section \ref{sec:con} concludes. 
\begin{figure*}[t!]
\includegraphics[width=\linewidth]{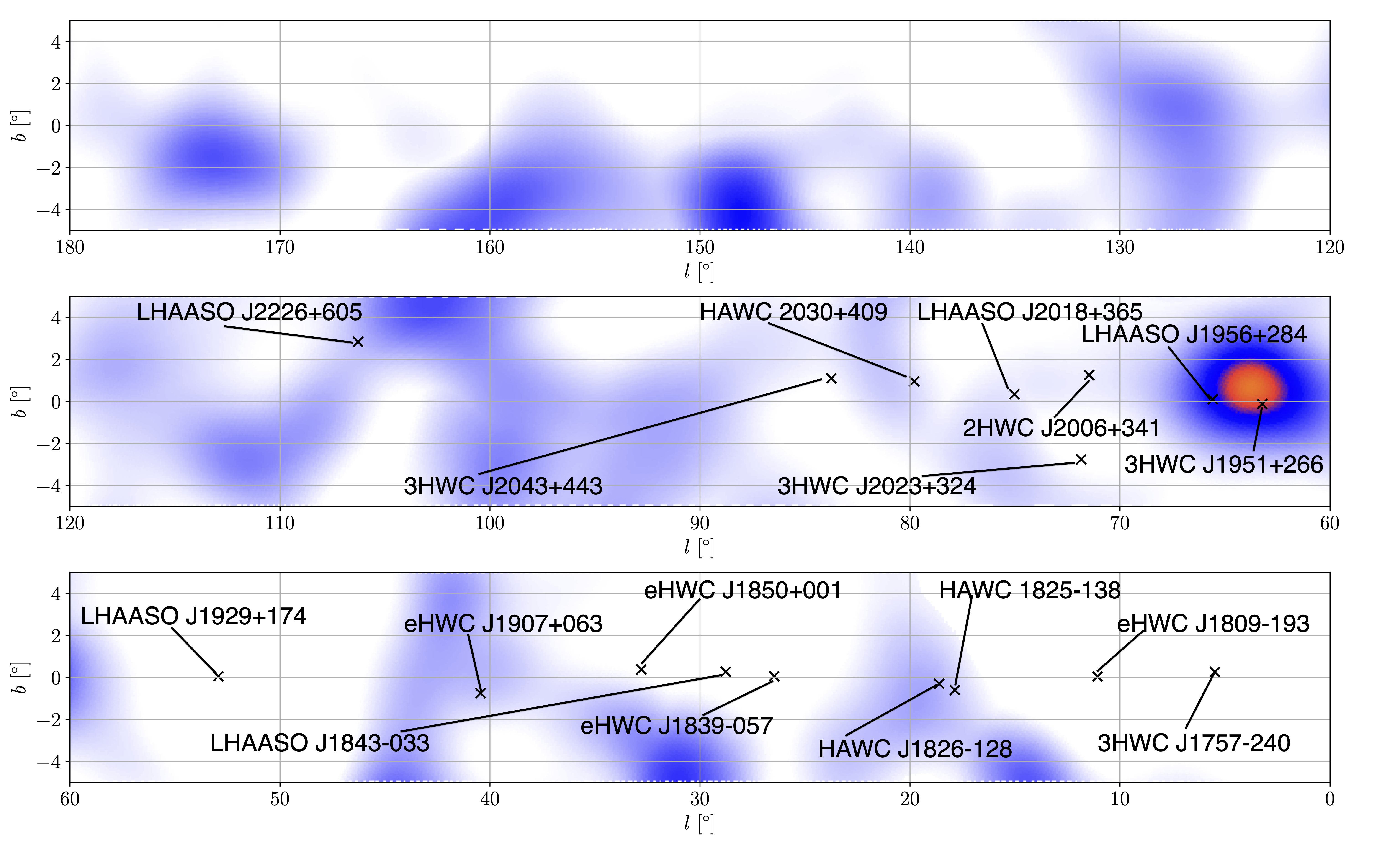}\\

\includegraphics[width=\linewidth]{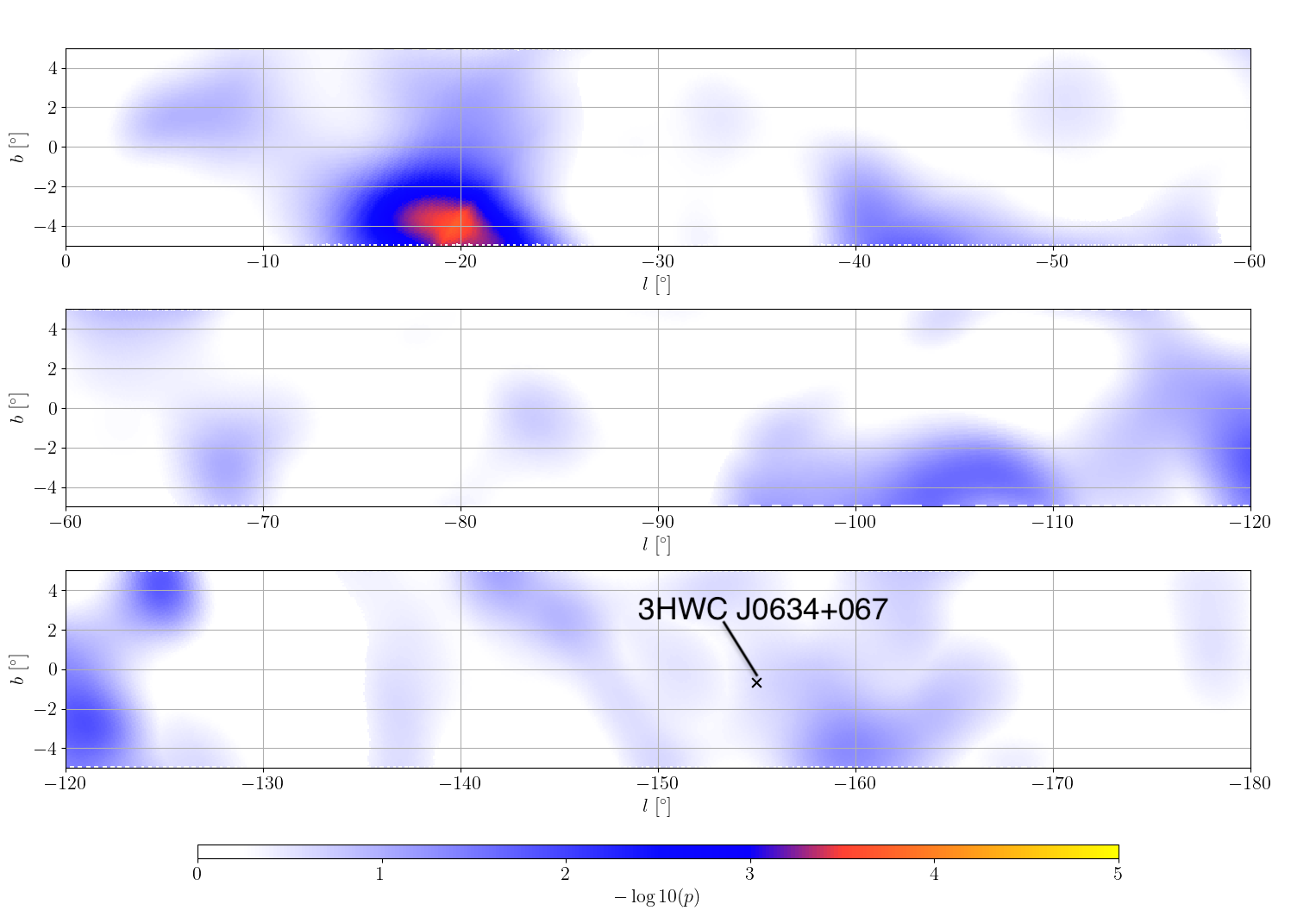}

 \caption{Local (pre-trials corrected) p-value map in Galactic coordinates for a $2.0^\circ$ source extension. The ROI locations used in the catalog search are labeled (see text for details). }
\label{fig:gal_map}
\end{figure*}


\section{\label{sec:ic}The IceCube Neutrino Observatory and Data}
The IceCube Neutrino Observatory is designed to detect cosmic neutrinos, most effectively above a few TeV in energy, via the Cherenkov radiation produced as a result of neutrino interactions in ice. The individual detector units, known as digital optical modules are embedded in Antarctic ice on 86 strings, forming a hexagonal array spanning a cubic kilometer of ice. Details about the detector and signal reconstruction can be found in \citep{icecube_detector,icecube_daq,Aartsen:2013vja}. 

Charged-current muon neutrino interactions produce muons that deposit energy in the detector in the form of tracks, which can be reconstructed with a directional accuracy of less than $1^\circ$ above 1 TeV. We use a data sample consisting of muon tracks collected between May 2011 and May 2020, with a total livetime of 3184 days. The sample has been used and validated in several searches for point sources with IceCube \citep{ic10yrs,IceCube:2021waz,IceCube:2022heu,IceCube:2022ccm}. The main background for astrophysical neutrinos in this data sample are track-like events from atmospheric neutrinos and muons produced during the interaction of cosmic rays with the atmosphere. 

In addition to track-like events, IceCube also detects showers or ``cascades" produced by neutral-current and (electron/tau neutrinos') charged current interactions. The direction of neutrino-induced cascades can be reconstructed with limited accuracy and have typical angular uncertainties of $\sim 15^\circ$.  While this work primarily uses tracks due to their superior sensitivity, we use cascades as a statistically independent data set to perform certain cross-checks (section \ref{sec:source_catalog_search}).

\begin{table*}
	\begin{center}
	\begin{tabular}{c|ccccccc}
	\toprule
	$\sigma$ [$^\circ$] & RA [$^\circ$]  & DEC [$^\circ$]  &  l [$^\circ$] & b [$^\circ$]  & $\hat{n}_s$ & $\hat{\gamma}$ & $\phi_{90\%}$ [$\mathrm{TeV}^{-1} \, \mathrm{cm}^{-2} \, \mathrm{s}^{-1}]$\\ 
	\hline
	$0.5$ & $296.98$   & $27.45$ & 63.53 & 1.00 & 80.3 & 3.10 & $5.13 \times 10^{-11}$ \\
	1.0 & $296.98$   & $27.45$ & 63.53 & 1.00 & 111.4 & 3.00 & $6.29 \times 10^{-11}$ \\
	$1.5$ & $297.42$   & $27.53$ &  63.80 & 0.71 & 150.5 & 3.03 & $6.61 \times 10^{-11}$ \\
	$2.0$ & $297.42$   & $27.53$ & 63.80 & 0.71 & 182.3 & 3.09 & $1.04 \times 10^{-10}$ \\
	\hline
	\end{tabular}
	\caption{Summary of results for the hottest spot in each scan along the Galactic plane for different extensions. The coordinates of the hottest spot, the best-fit number of signal events ($\hat{n}_s$), the spectral index ($\hat{\gamma}$) and the 90\% CL upper limit flux at 1~TeV are given for each search.} \label{tab:scan_upperlimits}
	\end{center}
\end{table*}

\section{\label{sec:ana}Analysis}
We use the unbinned maximum likelihood method to search for time-integrated excess neutrino emission above background from a given region in the sky as described in \citep{unbinned_llh}. The likelihood is formed by a product of probability densities over all events in the data,
\begin{equation}\label{eq:likelihood}
    \mathcal{L}(n_s,\gamma) = \prod_{i=1}^N \left[ \frac{n_s}{N} \mathcal{S}_i + \left(1-\frac{n_s}{N}\right)\mathcal{B}_i\right]
\end{equation}
where the fitted parameters are the number of signal events, $n_s$, and the spectral index, $\gamma$. $N$ is the number of total events in the data set, $\mathcal{S}_i$ is the signal probability density function (PDF), and $\mathcal{B}_i$ is the background PDF. The signal and the background PDFs contain a spatial term and an energy term. The computation of the PDFs are discussed in detail in \cite{icecube_7year}. Here we describe the two modifications that are used in this analysis to  focus on extended sources.

First, the spatial term in the signal PDF functionally depends on the extension of the source. The probability that the $i$th event came from an extended source at $\vec{x}_s$ is modeled by a 2D Gaussian given by,
\begin{equation}
    S_i = \frac{1}{2\pi \left(\sigma_i^2+\sigma_s^2\right)} \exp \left(- \frac{|\vec{x}_i-\vec{x}_s|^2}{2\left(\sigma_i^2+\sigma_s^2\right)} \right),
\end{equation}
 where $\vec{x}_i = \left(\alpha_i, \delta_i\right)$ is the $i$th event direction in right ascension and declination, $\sigma_i$ is the angular uncertainty of the $i$th event, and $\sigma_s$ is the source extension. 

The second modification is applied during the computation of $B_i$ to account for any signal contamination in the background. The background PDF at a given declination is calculated by randomizing the events in right ascension. Since the process uses actual data it may result in an overestimation of the background in the presence of a nearby source. The signal events from the source would be scrambled into the background. In order to avoid this contamination, we mask all regions that are potential sources of neutrino emission, before randomizing the right ascension of events. In this way, potential signal events are not included in the estimation of the background. An overall correction factor is applied to the background density to account for the fraction of sky that is masked during the calculation.

In this work, we mask out a disk of radius $1.5^\circ$ centered on the two known candidate sources of neutrinos: TXS 0506+056 and NGC 1068  \citep{txs,IceCube:2022der}. To account for any diffuse emission from the galaxy, we mask all events that have a Galactic latitude,  $|b| \leq 5^\circ$. The size is chosen based on the locations of Galactic TeV gamma-ray sources detected by HAWC and LHAASO, which are all within 5 degrees of the Galactic plane \citep{bib:3hwc_catalog,bib:lhaaso_12}.  We note that for the source extensions considered in this work ($\leq 2^\circ$), this analysis is sensitive to $\mathcal{O}(1\%)$ of the nominal Galactic diffuse flux measured in \citep{2023icgpdiffuse}.  For source extensions $< 5^\circ$, the inclusion of a model of Galactic plane emission in the background PDF has a negligible impact on the sensitivity.

Following the estimation of background, we maximize the likelihood in equation \ref{eq:likelihood} to determine the best-fit parameters, $n_s$ and $\gamma$, for a source with a fixed extension at a given location. In this work, we test for four different source extensions: $0.5^\circ$, $1.0^\circ$, $1.5^\circ$ and $2.0^\circ$. The different locations searched are described below.

\subsection{The Galactic Plane Scan}
The first search consists of a largely model-independent scan of every location in the Galactic plane between $-5^\circ \leq b \leq 5^\circ$, making no assumptions about the detailed morphology or spectral slope of the underlying emission. The galactic latitude cut is based on the measurements of diffuse TeV gamma-ray emission from the plane \citep{tibetasgcollaborationFirstDetectionSubPeV2021,LHAASO:2023gne}. We bin the sky into equal-area HEALpix  pixels with the mean spacing between pixels set to $0.115^\circ$ \citep{2005ApJ...622..759G}. At each pixel location, we fit for a neutrino source of a fixed extension and a spectrum described by a simple power law with spectral index $\gamma$. The total number of signal events from the source and $\gamma$ are the free parameters of the fit, which determine the differential flux at 1 TeV (reference energy). For each location, we have four sets of fits corresponding to the four source extensions. The test statistic for a fit is given by,
\begin{equation}
    \text{TS} = 2 \log \left( \frac{\mathcal{L}\left(\hat{n}_s,\hat{\gamma}\right)}{\mathcal{L}\left(n_s=0\right)}  \right),
\end{equation}
where $\hat{n}_s$ and $\hat{\gamma}$ are the best-fit values of the free parameters. For each fit, a local or pre-trials p-value is determined by comparing the observed TS with a TS distribution from an ensemble of background-only trials. Since a very large number of locations are tested multiple times for possible neutrino emission, a further ``trials correction" factor for the lowest p-value is calculated by simulating the whole search 5000 times on background-only data. 

No significant emission from a source with an extension between $0.5^\circ$ and $2^\circ$ is observed at any location in the Galactic plane. Figure \ref{fig:gal_map} shows the local p-value map of the Galactic plane assuming a source extension of $2^\circ$. The upper limits on the flux for the location with the lowest p-value are shown in table \ref{tab:scan_upperlimits}. 

\begin{table*}[htb]
	\begin{center}
	\resizebox{0.95\textwidth}{!}{
	\begin{tabular}{c|ccccc}
	\hline
	\multirow{2}*{Region of Interest} & \multirow{2}*{RA [${}^\circ$]} & \multirow{2}*{Dec [${}^\circ$]} & \multirow{2}*{$l$ [${}^\circ$]} & \multirow{2}*{$b$ [${}^\circ$]} & Possible Sources \\ 
	  &  &  & & & and Associated Extension \\ 
	\hline
	ROI-1  & 95.32  & 38.21 & 175.44 & 10.97  & 3HWC J0621+382 ($0.5^\circ$) \citep{bib:3hwc_catalog} \\
	\hline
	ROI-2  & 95.47  & 37.92 & 175.76 & 10.95 & LHAASO J0621+3755 ($0.4^\circ$) \citep{bib:LHAASO_extended} \\
	\hline
	ROI-3  & 98.66  &  6.73  & 205.03 & -0.65 & 3HWC J0634+067 ($0.5^\circ$) \citep{bib:3hwc_catalog} \\
	\hline
	ROI-4  & 269.3  & -24.09 & 5.49 & 0.25 & 3HWC J1757-240 ($1.0^\circ$) \citep{bib:3hwc_catalog} \\
	\hline
	ROI-5  & 272.46 & -19.34& 11.07 & 0.03 & eHWC J1809-193 ($0.34^\circ$) \citep{bib:astronomy_HAWC_GalacticPlane56TeV} \\
	\hline
    \multirow{2}*{ROI-6} & \multirow{2}*{276.42} & \multirow{2}*{-13.66} & \multirow{2}*{17.87} & \multirow{2}*{-0.61}& HAWC J1825-138 ($0.47^\circ$) \cite{bib:hawc_j1825} \\ 
	  &  &  & && LHAASO J1825-1326 ($0.3^\circ$) \citep{bib:lhaaso_12} \\
	\hline
	ROI-7  & 276.5  & -12.86  & 18.61 & -0.31 & HAWC J1826-128 ($0.2^\circ$) \citep{bib:hawc_j1825} \\
	\hline
    \multirow{2}*{ROI-8} & \multirow{2}*{279.86} & \multirow{2}*{-5.73} & \multirow{2}*{26.47} & \multirow{2}*{0.05} & eHWC J1839-057 ($0.34^\circ$) \cite{bib:astronomy_HAWC_GalacticPlane56TeV} \\ 
	  &  &  & &&LHAASO J1839-0545 ($0.3^\circ$) \citep{bib:lhaaso_12} \\
	\hline
    \multirow{2}*{ROI-9} & \multirow{2}*{280.73} & \multirow{2}*{-3.58} & \multirow{2}*{28.78} & \multirow{2}*{0.26} & eHWC J1842-035 ($0.39^\circ$) \citep{bib:astronomy_HAWC_GalacticPlane56TeV} \\ 
	  &  &  &&& LHAASO J1843-0338 ($0.3^\circ$) \citep{bib:lhaaso_12} \\
	\hline
    \multirow{2}*{ROI-10} & \multirow{2}*{282.47} & \multirow{2}*{0.05} & \multirow{2}*{32.80} & \multirow{2}*{0.37} & eHWC J1850+001 ($0.37^\circ$) \citep{bib:astronomy_HAWC_GalacticPlane56TeV} \\ 
	  &  &  &&& LHAASO J1849-0003 ($0.3^\circ$) \cite{bib:lhaaso_12} \\
	\hline
    \multirow{2}*{ROI-11} & \multirow{2}*{286.98} & \multirow{2}*{6.34} & \multirow{2}*{40.45} & \multirow{2}*{-0.76} & eHWC J1907+063 ($0.67^\circ$) \citep{bib:astronomy_HAWC_GalacticPlane56TeV} \\ 
	  &  &  &&& LHAASO J1908+0621 ($0.3^\circ$) \citep{bib:lhaaso_12} \\
	\hline
	ROI-12  & 292.25  & 17.75 & 52.94 & 0.04  & LHAASO J1929+1745  ($0.3^\circ$) \citep{bib:lhaaso_12} \\
	\hline
	ROI-13 & 297.9  & 26.61 & 63.23 & -0.1  & 3HWC J1951+266 ($0.5^\circ$) \citep{bib:3hwc_catalog} \\
	\hline
	ROI-14  & 299.05  & 28.75  & 65.58 & 0.10 & LHAASO J1956+2845  ($0.3^\circ$) \citep{bib:lhaaso_12} \\
	\hline
	ROI-15 & 301.55  & 34.35& 71.46 & 1.25 & 2HWC J2006+341 ($0.72^\circ$) \citep{bib:hawc_j2006} \\
	\hline
	  &  &  &&& eHWC J2019+368 ($0.3^\circ$) \citep{bib:astronomy_HAWC_GalacticPlane56TeV} \\
    ROI-16  & 304.90  & 36.82 & 75.03 & 0.34 & LHAASO J2018+3651 ($0.3^\circ$) \cite{bib:lhaaso_12} \\
	  &  &  &&& TASG J2019+368 ($0.28^\circ$) \citep{bib:tibet_cygnus} \\
	\hline
	ROI-17  & 305.81 & 32.44 & 71.85 & -2.77 & 3HWC J2023+324 ($1.0^\circ$) \citep{bib:3hwc_catalog}\\
	\hline
	  &  &  &&& eHWC J2030+412 ($0.18^\circ$) \citep{bib:astronomy_HAWC_GalacticPlane56TeV} \\
    ROI-18  & 307.81  & 41.07 & 79.80 & 0.95 & LHAASO J2032+4102 ($0.3^\circ$) \citep{bib:lhaaso_12} \\
	  &  &  &&& HAWC J2030+409  ($2.13^\circ$) \citep{bib:hawc_cygnus} \\
	\hline
	ROI-19  & 310.89 & 44.3 & 83.74 & 1.10 & 3HWC J2043+443 ($0.5^\circ$) \citep{bib:3hwc_catalog}\\
	\hline
	ROI-20  & 336.75  & 60.95 & 106.28 & 2.84  & LHAASO J2226+6057  ($0.3^\circ$) \citep{bib:lhaaso_12} \\
	\hline
	\end{tabular}
	}
	\caption{The locations of the ROI and the possible sources therein that are used in the catalog search.} 
	\label{tab:ROI}
	\end{center}
\end{table*}

\subsection{\label{sec:source_catalog_search}The Catalog Search}
The second search focuses on neutrino emission from known extended sources of TeV gamma-ray emission. A targeted catalog search has the advantage of using multimessenger information to pin down potential sources, resulting in a reduced trials factor compared to the all-sky search. For this analysis, we select a catalog of sources that exhibit an extended morphology as observed by TeV gamma-ray observatories \citep{2008ICRC....3.1341W}. The sources that pass this criteria are labeled in figure \ref{fig:gal_map}. In some cases, two or more reported sources are possibly associated and are less than $0.5^\circ$ of each other. We group these sources into ROI and choose a location equidistant from all sources as the central location of the ROI. Isolated sources are labeled as individual ROI. This procedure gives us a catalog of 20 ROIs to search for neutrino emission with an extension between $0.5^\circ$ and $2.0^\circ$. Table \ref{tab:ROI} lists the ROI locations and the corresponding sources.
\begin{figure}[h!]

         \includegraphics[width=\linewidth]{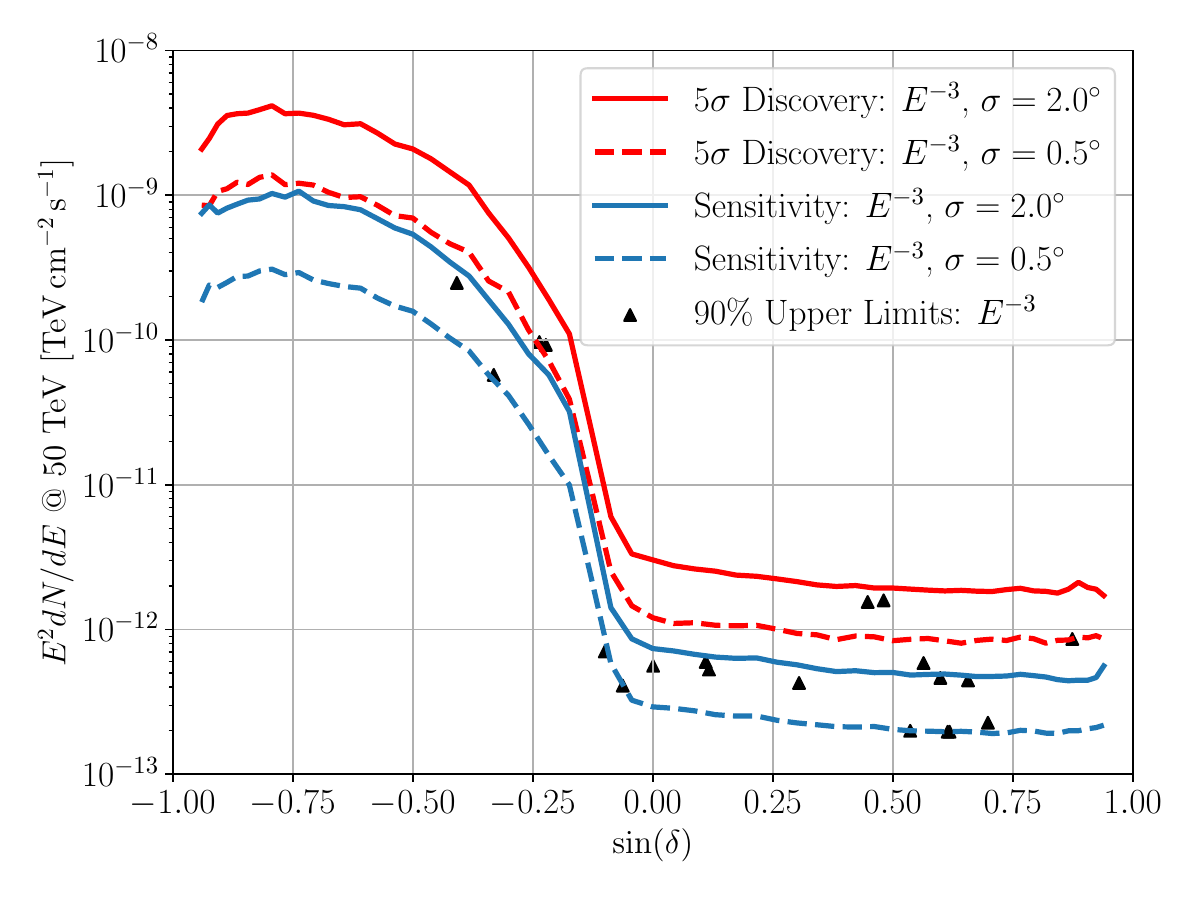}  
 \caption{The 90\% CL limits on the neutrino flux at 50 TeV from the ROIs in the catalog search, assuming a spectral index of 3. The solid red and blue lines show the 5$\sigma$ discovery potential and sensitivity for a source with  $\sigma_s = 2.0^\circ$. The dashed red and blue lines show the 5$\sigma$ discovery potential and sensitivity for a source with an extension of $0.5^\circ$. See text for the definitions of the ROIs.}
\label{fig:limits}
\end{figure}

For each ROI, we fit for $n_s$ and $\gamma$ for extensions $0.5^\circ,1.0^\circ,1.5^\circ$ , and $2.0^\circ$ as described above. No significant extended emission is observed in any of the ROIs resulting in constraints on the total neutrino flux from each region. Table \ref{tab:ROIlims} provides the 90\% upper limits on the differential neutrino flux from each ROI at a reference energy of 50 TeV. See the appendix for detailed fit results. For each ROI, we provide the upper limit corresponding to the extension that gives the smallest p-value during the various fits, for $\gamma = 3$.  

We also compare our constraints to the expected muon neutrino flux, $\phi_{\nu} (E_{\nu})$, from the sources within each ROI.  Following the methods in \cite{Ahlers:2013xia}, we calculate $\phi_{\nu} (E_{\nu})$  under the assumption that all of the observed gamma-ray flux, $\phi_{\gamma}(E_{\gamma})$ from a given source is produced in $pp$ collisions, and is therefore accompanied by neutrinos. We only consider $pp$ interactions here, since those are expected to dominate over p$\gamma$ interaction in the Galactic plane region \citep{Murase:2013rfa,Ahlers:2013xia}. $\phi_{\nu} (E_{\nu}) $ is then given by,   $\phi_{\nu} (E_{\nu}) = 2^{1-\gamma}\phi_{\gamma}(E_{\gamma})$, where $\gamma$ is the common spectral index of the neutrino and gamma-ray emission, and the neutrino energy $E_{\nu}$ is half the gamma-ray energy, $E_\gamma$ \citep{Ahlers:2013xia}.   

The ROI considered in this work include notable PeVatron candidates. For instance, we obtain the strongest limits in terms of constraining the hadronic emission from ROI-18, with $\phi_{90\%}/\phi_{\nu}$ of $\sim 0.5$, where $\phi_{\nu}$ is the predicted neutrino flux assuming all gamma rays are hadronic. This ROI is part of the Cygnus region and includes HAWC J2030+409, LHAASO J2032+4102 and eHWC J2030+412 \citep{bib:hawc_cygnus,bib:tibet_cygnus}. ROI-20 is co-located with LHAASO J2226+6057, which is $0.14^\circ$ away from the SNR G106.3+02.7 (also associated with HAWC J2227+610) \citep{HAWC:2020nvc}, which is another proposed hadronic accelerator \citep{Fang:2022uge}. In this region, our most conservative upper limit is a factor of $\sim 2.7$ above the hadronic scenario, implying the need for improved sensitivity to detect neutrinos from this potential cosmic-ray accelerator.

Figure \ref{fig:limits} shows the upper limits on the flux from each ROI for the extension with the highest TS assuming $\gamma = 3$. Also shown are the sensitivity and discovery potential as a function of source declination.

\begin{figure}[h!]

\includegraphics[width=\linewidth]{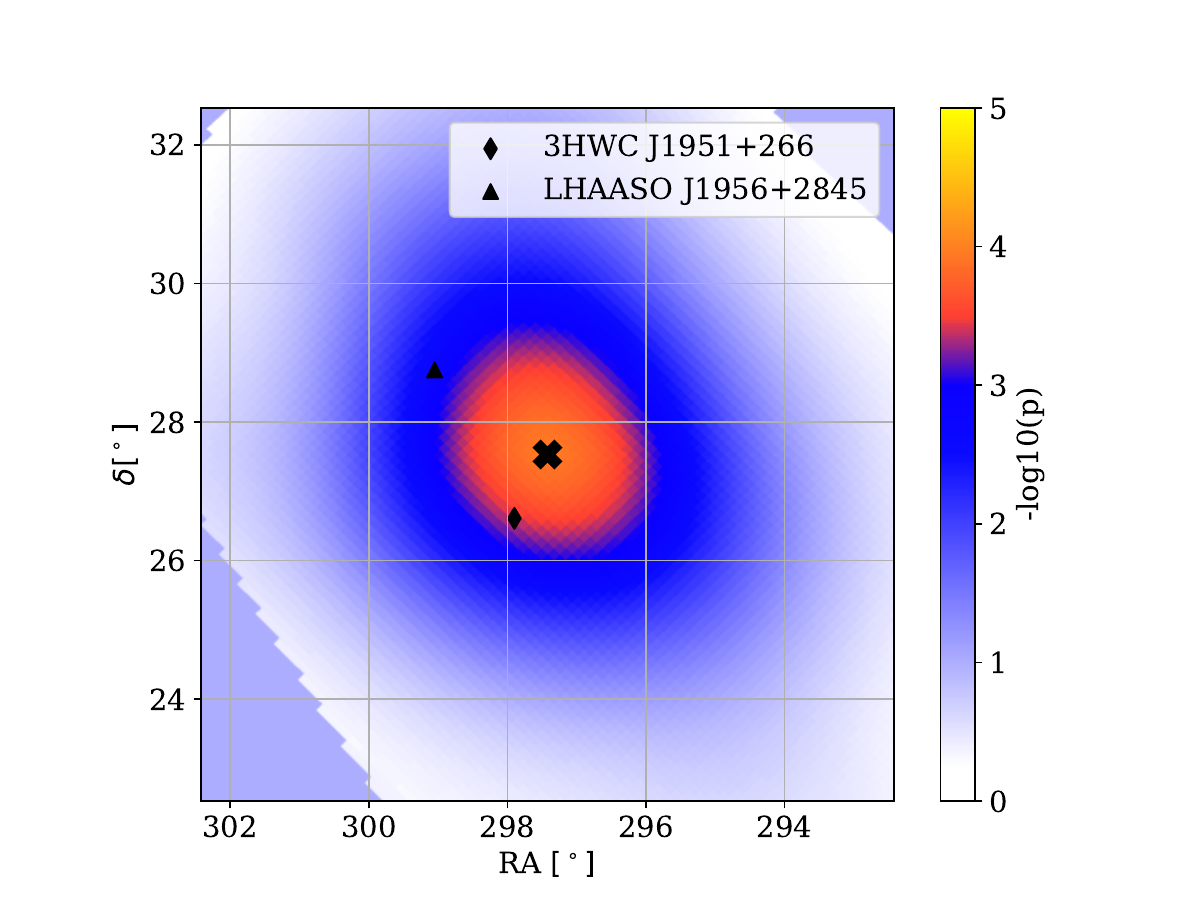}  
 \caption{The region of the Galactic plane with the lowest
p-values in the general scan as well as the catalog search.
The general hot spot is marked with a cross. Sources
corresponding to ROI-13 (3HWC J1951+266) and ROI-14
(LHAASO J1956+2845) from the catalog search are also
labeled. The map shows pre-trials corrected p-values only.}
\label{fig:hotpsot}
\end{figure}

\subsection{The Most Significant Region}
The highest TS in the catalog search is obtained for ROI-13 at the location of 3HWC J1951+266 for an extension of $1.5^\circ$, with a best-fit flux of $5.2 \times 10^{-13}$ TeV cm$^{-2}$ s$^{-1}$ at 100 TeV and $\gamma = $  3.03.  For this ROI, we perform a scan across a finer grid of extensions to determine the source extension that best describes the potential neutrino signal. The local significance is further corrected for multiple testing (including the 20 ROIs and several extensions) by performing all the tests on 5000 simulations and constructing a background-only p-value distribution. The global (or trials-corrected) p-value is then given by the probability of obtaining a particular local p-value of ROI-13 in the aforementioned distribution. We obtain the lowest p-value for an extension of $1.7^\circ$ at a global significance of $2.6\sigma$.  The hottest spot in the Galactic plane scan is located $1.02^\circ$ away from ROI-13 and $1.88^\circ$ away from ROI-14. Figure \ref{fig:hotpsot} shows the most significant locations in both the catalog search and the general scan for an extension of $1.5^\circ$. Following this result, we also study the location of the hotspot using an independent dataset of neutrino-induced cascades and find the best-fit flux and spectral index to be consistent with the tracks' results. However, the result is not significant enough to qualify as evidence for emission. 

\begin{table*}
	\begin{center}
	\resizebox{0.90\textwidth}{!}{
	\begin{tabular}{c|cccc}
	\hline
	Region of & Gamma-Ray & $\phi_\nu$ from $pp$ Collisions & $\phi_{90\%}$ at 50 TeV & \multirow{2}*{$\frac{\phi_{90\%}}{\phi_\nu}$}\\ 
	Interest &   Source Name  & $\left(\mathrm{TeV}^{-1}\,\mathrm{cm}^{-2}\,\mathrm{s}^{-1}\right)$ & $\left(\mathrm{TeV}^{-1}\mathrm{cm}^{-2}\mathrm{s}^{-1}\right)$ & \\
	\hline
	ROI-18 & HAWC J2030+409    & $3.88\times10^{-16}$ & $1.78\times10^{-16}$  & $0.459$ \\
    \hline
    ROI-11 & eHWC J1907+063    & $4.95\times10^{-16}$ & $2.39\times10^{-16}$  & $0.482$ \\ 
    \hline
	ROI-16 & eHWC J2019+368    & $3.82\times10^{-16}$ & $1.85\times10^{-16}$  & $0.485$  \\
    \hline
    ROI-9  & eHWC J1842-035    & $3.04\times10^{-16}$ & $1.64\times10^{-16}$  & $0.540$  \\ 
    \hline
	ROI-7  & HAWC J1826-128    & $5.54\times10^{-14}$ & $3.70\times10^{-14}$  & $0.668$ \\
	\hline 
    ROI-8  & eHWC J1839-057    & $3.04\times10^{-16}$ & $2.82\times10^{-16}$  & $0.928$ \\
    \hline
	ROI-18 & eHWC J2030+412    & $1.82\times10^{-16}$ & $1.78\times10^{-16}$  & $0.978$  \\
    \hline
    ROI-10 & eHWC J1850+001    & $2.23\times10^{-16}$ & $2.25\times10^{-16}$  & $1.01$ \\ 
    \hline
	ROI-16 & TASG J2019+368    & $1.79\times10^{-16}$ & $1.85\times10^{-16}$  & $1.04$  \\
    \hline
	ROI-2  & LHAASO J0621+3755 & $4.28\times10^{-17}$ & $5.79\times10^{-17}$  & $1.35$ \\
    \hline
	ROI-11 & LHAASO J1908+0621 & $1.66\times10^{-16}$ & $2.39\times10^{-16}$  & $1.44$ \\
    \hline
	ROI-9  & LHAASO J1843-0338 & $8.91\times10^{-17}$ & $1.64\times10^{-16}$  & $1.84$  \\
    \hline
	ROI-1  & 3HWC J0621+382    & $2.93\times10^{-17}$ & $5.64\times10^{-17}$  & $1.92$ \\
	\hline
    ROI-6  & HAWC J1825-138    & $1.80\times10^{-14}$ & $3.87\times10^{-14}$  & $2.16$  \\ 
    \hline
	ROI-19 & 3HWC J2043+443    & $3.95\times10^{-17}$ & $9.08\times10^{-17}$  & $2.30$ \\
    \hline
	ROI-10 & LHAASO J1849-0003 & $9.03\times10^{-17}$ & $2.25\times10^{-16}$  & $2.49$  \\
    \hline
	ROI-20 & LHAASO J2226+6057 & $1.28\times10^{-16}$ & $3.44\times10^{-16}$  & $2.69$  \\
    \hline
    ROI-18 & LHAASO J2032+4102 & $6.59\times10^{-17}$ & $1.78\times10^{-16}$  & $2.70$ \\
    \hline
    ROI-16 & LHAASO J2018+3651 & $6.10\times10^{-17}$ & $1.85\times10^{-16}$  & $3.03$ \\
    \hline
	ROI-17 & 3HWC J2023+324    & $2.10\times10^{-17}$ & $6.62\times10^{-17}$  & $3.15$  \\
    \hline
	ROI-8  & LHAASO J1839-0545 & $8.54\times10^{-17}$ & $2.82\times10^{-16}$  & $3.30$  \\
    \hline
	ROI-12 & LHAASO J1929+1745 & $4.64\times10^{-17}$ & $1.71\times10^{-16}$  & $3.69$ \\
    \hline
	ROI-3  & 3HWC J0634+067    & $4.30\times10^{-17}$ & $2.12\times10^{-16}$  & $4.92$  \\
	\hline
	ROI-14 & LHAASO J1956+2845 & $5.00\times10^{-17}$ & $6.36\times10^{-16}$  & $12.7$  \\
    \hline
	ROI-13 & 3HWC J1951+266    & $3.20\times10^{-17}$ & $6.20\times10^{-16}$  & $19.4$  \\
    \hline
	ROI-5  & eHWC J1809-193    & $4.86\times10^{-16}$ & $2.17\times10^{-14}$  & $44.7$ \\
    \hline
	ROI-6  & LHAASO J1825-1326 & $4.36\times10^{-16}$ & $3.87\times10^{-14}$  & $88.9$  \\
    \hline
	ROI-15 & 2HWC J2006+341    & $8.37\times10^{-19}$ & $2.35\times10^{-16}$  & $280$  \\
    \hline
	ROI-4  & 3HWC J1757-240    & $8.41\times10^{-17}$ & $9.93\times10^{-14}$  & $1.18\times10^{3}$ \\
	\hline
	\end{tabular}
	}
	\caption{Catalog search results in order of most constraining to least constraining. For each
ROI, we show the
expected neutrino flux at 50 TeV assuming
hadronic origins of the associated gamma-ray
emission, the 90\% CL limits assuming $\gamma = 3.0$, and the ratio of the upper
limit to the expected neutrino emission.} 
	\label{tab:ROIlims}
	\end{center}
\end{table*}

\section{\label{sec:con}Conclusions}
We perform a targeted search for spatially extended neutrino emission in the Milky Way utilizing ten years of neutrino track-like events in IceCube. We focus on potential source extensions between $0.5^\circ$ and $2.0^\circ$ in a general scan across the Galactic plane and a catalog search with extended regions of TeV gamma-ray sources. The most significant location is a $1.7^\circ$ region centered on the unidentified source 3HWC J1951+266 and is found to be inconsistent with the background-only hypothesis at 2.6$\sigma$ after trials correction. We emphasize that this is still below our threshold for evidence of significant emission. Our analysis also places constraints on neutrino emisson from a number of regions hypothesized to contain PeVatron candidates including the Cygnus cocoon, and the Boomerang supernova remnant. 

We encourage further multiwavelength campaigns across the Galactic plane in light of these new constraints. Such  studies would complement IceCube observations in helping understand the emission mechanisms underlying various regions in the plane. Furthermore, a large fraction of the Galactic plane lies in the Southern sky, where IceCube has limited sensitivity to individual sources, and TeV gamma-ray surveys have limited coverage.  In the coming years, more data with IceCube, as well as a number of near-future and planned observatories like KM3Net \citep{KM3NeT:2018wnd}, IceCube Gen-2 \citep{IceCube-Gen2:2020qha}, P-ONE \citep{P-ONE:2020ljt} and Baikal-GVD \citep{Baikal-GVD:2021ypx} will be able to probe the Galactic plane for PeVatrons in detail with better coverage and improved angular resolution.

\clearpage
\appendix

\begin{figure}
	\begin{center}
	\includegraphics[width=0.5\linewidth]{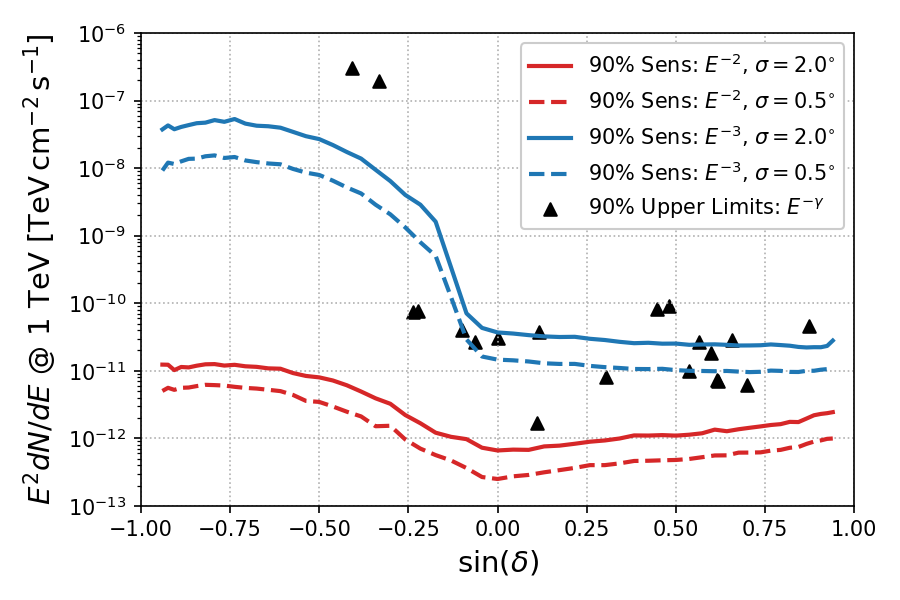}
	\caption{The 90\% upper-limit fluxes at a pivot energy of 1~TeV for the source extensions with the smallest pre-trial p-value for each ROI, shown as the black triangles. The upper limits are compared to the 90\% CL sensitivity curves with $\sigma_s = 0.5^\circ$ (dashed lines) and $\sigma_s = 2.0^\circ$ (solid lines) at $\gamma = 2.0$ (red lines) and $\gamma = 3.0$  (blue lines).} \label{fig:catalog_max_limits}
	\end{center}
\end{figure}

Here we provide additional detailed results of the fits. Table \ref{tab:catalog_post_results_ROI13} reports the summary of various fits for ROI-13. Table \ref{tab:catalog_max_fluxes} reports the fit results and upper limits for all ROI. Figure \ref{fig:catalog_max_limits} shows the upper limits for all ROI, and the sensitivity of the analysis for two different assumed spectral indices.  

\begin{table}[h]
	\begin{center}
	\resizebox{0.6\textwidth}{!}{
	\begin{tabular}{c|cccc}
	\hline
	Extension (${}^\circ$) & $\hat{n}_s$ & $\hat{\gamma}$ & $\mathrm{p}_{\mathrm{pre}}$ ($\sigma_{\mathrm{pre}}$) \\ 
	\hline
	$1.0^\circ$ & 99.33  & 3.03 & $3.87 \times 10^{-4}$ ($3.36\sigma$)  \\
	$1.1^\circ$ & 108.27 & 3.03 & $3.80 \times 10^{-4}$ ($3.37\sigma$)  \\
	$1.2^\circ$ & 116.95 & 3.04 & $2.67 \times 10^{-4}$ ($3.46\sigma$)  \\
	$1.3^\circ$ & 125.43 & 3.05 & $2.07 \times 10^{-4}$ ($3.53\sigma$)  \\
	$1.4^\circ$ & 133.63 & 3.06 & $2.07 \times 10^{-4}$ ($3.53\sigma$)  \\
	$1.5^\circ$ & 141.52 & 3.07 & $1.40 \times 10^{-4}$ ($3.63\sigma$)  \\
	$1.6^\circ$ & 149.03 & 3.08 & $1.73 \times 10^{-4}$ ($3.58\sigma$)  \\
	\boldmath $1.7^\circ$ & \boldmath $156.12$ & \boldmath $3.10$ & \boldmath $1.27 \times 10^{-4}$ ($3.66\sigma$)  \\
	$1.8^\circ$ & 163.06 & 3.12 & $2.27 \times 10^{-4}$ ($3.51\sigma$)  \\
	$1.9^\circ$ & 169.44 & 3.13 & $1.67 \times 10^{-4}$ ($3.59\sigma$)  \\
	$2.0^\circ$ & 175.29 & 3.14 & $2.53 \times 10^{-4}$ ($3.48\sigma$)  \\
	\hline
	
	Extension (${}^\circ$) & $\hat{n}_s$ & $\hat{\gamma}$ & $\mathrm{p}_{\mathrm{post}}$ ($\sigma_{\mathrm{post}}$) \\ 
	\hline
    $1.7^\circ$ & $156.12$ & $3.10$ & $4.50 \times 10^{-3}$ ($2.61\sigma$)  \\
	\hline
	\end{tabular}
	}
	\caption{Results for each source extension evaluated at the location of ROI-13: (RA,DEC)=$(297.9^\circ,26.61^\circ)$ (top). The observed number of signal events, $\hat{n}_s$ and the spectral index, $\hat{\gamma}$, are also reported. A post-trial p-value was obtained for the hottest extension ($1.7^\circ$) by taking into account all 20 ROI locations and 16 possible extensions in the finer scan, ranging from $0.5^\circ$ to $2.0^\circ$ in steps of $0.1^\circ$.} \label{tab:catalog_post_results_ROI13}
	\end{center}
\end{table}


\begin{table}[h]
	\begin{center}
	\resizebox{0.8\textwidth}{!}{
	\begin{tabular}{c|cccc}
	\hline
	\multirow{2}*{Region of Interest} & \multirow{2}*{Extension ($^\circ$)} & \multirow{2}*{$\hat{n}_s$} & \multirow{2}*{$\hat{\gamma}$} & $\phi_{90\%}$ at 50 TeV \\  
	 & & & & $(\mathrm{TeV}^{-1}\mathrm{cm}^{-2}\mathrm{s}^{-1})$\\ 
	\hline
    ROI-1  & $0.5^\circ$ & 0.0   & 3.00 & $5.64\times 10^{-17}$ \\
    ROI-2  & $0.5^\circ$ & 0.0   & 3.00 & $5.79\times 10^{-17}$ \\
    ROI-3  & $0.5^\circ$ & 32.9  & 3.56 & $3.39\times 10^{-17}$ \\
    ROI-4  & $1.0^\circ$ & 14.1  & 3.54 & $2.92\times 10^{-13}$ \\
    ROI-5  & $0.5^\circ$ & 0.0   & 3.75 & $8.19\times 10^{-14}$ \\
    ROI-6  & $1.5^\circ$ & 7.0   & 2.39 & $6.54\times 10^{-15}$ \\
    ROI-7  & $1.5^\circ$ & 9.2   & 2.40 & $6.48\times 10^{-15}$ \\
    ROI-8  & $0.5^\circ$ & 9.3   & 3.08 & $2.34\times 10^{-16}$ \\
    ROI-9  & $0.5^\circ$ & 0.6   & 4.00 & $4.30\times 10^{-18}$ \\
    ROI-10 & $2.0^\circ$ & 18.9  & 3.07 & $1.82\times 10^{-16}$ \\
    ROI-11 & $0.5^\circ$ & 7.8   & 2.14 & $3.90\times 10^{-16}$ \\
    ROI-12 & $0.5^\circ$ & 18.6  & 2.54 & $3.87\times 10^{-16}$ \\
    ROI-13 & $1.5^\circ$ & 141.5 & 3.07 & $4.91\times 10^{-16}$ \\
    ROI-14 & $2.0^\circ$ & 149.5 & 3.18 & $3.59\times 10^{-16}$ \\
    ROI-15 & $2.0^\circ$ & 31.2  & 4.00 & $4.23\times 10^{-18}$ \\
    ROI-16 & $0.5^\circ$ & 24.7  & 2.83 & $2.76\times 10^{-16}$ \\
    ROI-17 & $0.5^\circ$ & 1.9   & 3.21 & $3.51\times 10^{-17}$ \\
    ROI-18 & $0.5^\circ$ & 30.6  & 3.52 & $2.93\times 10^{-17}$ \\
    ROI-19 & $0.5^\circ$ & 6.5   & 2.63 & $2.05\times 10^{-16}$ \\
    ROI-20 & $2.0^\circ$ & 85.0  & 3.38 & $8.11\times 10^{-17}$ \\
	\hline
	\end{tabular}
	}
	\caption{The 90\% upper-limit fluxes at a pivot energy of 50~TeV for the source extensions with the smallest pre-trial p-value for each ROI. The associated extension, fitted number of signal events, $\hat{n}_s$, and the fitted spectral index, $\hat{\gamma}$ are shown.} \label{tab:catalog_max_fluxes}
	\end{center}
\end{table}


\clearpage

\section*{Acknowledgements}
The IceCube Collaboration acknowledges the significant contributions to this manuscript from Devyn Rysewyk and Mehr Un Nisa. We also acknowledge support from: USA {\textendash} U.S. National Science Foundation-Office of Polar Programs,
U.S. National Science Foundation-Physics Division,
U.S. National Science Foundation-EPSCoR,
Wisconsin Alumni Research Foundation,
Center for High Throughput Computing (CHTC) at the University of Wisconsin{\textendash}Madison,
Open Science Grid (OSG),
Advanced Cyberinfrastructure Coordination Ecosystem: Services {\&} Support (ACCESS),
Frontera computing project at the Texas Advanced Computing Center,
U.S. Department of Energy-National Energy Research Scientific Computing Center,
Particle astrophysics research computing center at the University of Maryland,
Institute for Cyber-Enabled Research at Michigan State University,
and Astroparticle physics computational facility at Marquette University;
Belgium {\textendash} Funds for Scientific Research (FRS-FNRS and FWO),
FWO Odysseus and Big Science programmes,
and Belgian Federal Science Policy Office (Belspo);
Germany {\textendash} Bundesministerium f{\"u}r Bildung und Forschung (BMBF),
Deutsche Forschungsgemeinschaft (DFG),
Helmholtz Alliance for Astroparticle Physics (HAP),
Initiative and Networking Fund of the Helmholtz Association,
Deutsches Elektronen Synchrotron (DESY),
and High Performance Computing cluster of the RWTH Aachen;
Sweden {\textendash} Swedish Research Council,
Swedish Polar Research Secretariat,
Swedish National Infrastructure for Computing (SNIC),
and Knut and Alice Wallenberg Foundation;
European Union {\textendash} EGI Advanced Computing for research;
Australia {\textendash} Australian Research Council;
Canada {\textendash} Natural Sciences and Engineering Research Council of Canada,
Calcul Qu{\'e}bec, Compute Ontario, Canada Foundation for Innovation, WestGrid, and Compute Canada;
Denmark {\textendash} Villum Fonden, Carlsberg Foundation, and European Commission;
New Zealand {\textendash} Marsden Fund;
Japan {\textendash} Japan Society for Promotion of Science (JSPS)
and Institute for Global Prominent Research (IGPR) of Chiba University;
Korea {\textendash} National Research Foundation of Korea (NRF);
Switzerland {\textendash} Swiss National Science Foundation (SNSF);
United Kingdom {\textendash} Department of Physics, University of Oxford.
\bibliography{mbib}{}

\begin{thebibliography}{}
\expandafter\ifx\csname natexlab\endcsname\relax\def\natexlab#1{#1}\fi
\providecommand{\url}[1]{\href{#1}{#1}}
\providecommand{\dodoi}[1]{doi:~\href{http://doi.org/#1}{\nolinkurl{#1}}}
\providecommand{\doeprint}[1]{\href{http://ascl.net/#1}{\nolinkurl{http://ascl.net/#1}}}
\providecommand{\doarXiv}[1]{\href{https://arxiv.org/abs/#1}{\nolinkurl{https://arxiv.org/abs/#1}}}

\bibitem[{Aartsen {et~al.}(2018{\natexlab{a}})Aartsen, Ackermann, Adams,
  Aguilar, Ahlers, Ahrens, Al~Samarai, Altmann, Andeen, \& et~al.}]{txs}
Aartsen, M., Ackermann, M., Adams, J., {et~al.} 2018{\natexlab{a}}, Science,
  361, 147–151, \dodoi{10.1126/science.aat2890}

\bibitem[{Aartsen {et~al.}(2018{\natexlab{b}})Aartsen, Ackermann, Adams,
  Aguilar, Ahlers, Ahrens, Al~Samarai, Altmann, Andeen, \& et~al.}]{txs2}
---. 2018{\natexlab{b}}, Science, 361, eaat1378,
  \dodoi{10.1126/science.aat1378}

\bibitem[{Aartsen {et~al.}(2014{\natexlab{a}})}]{IceCube:2014vjc}
Aartsen, M.~G., {et~al.} 2014{\natexlab{a}}, Astrophys. J., 796, 109,
  \dodoi{10.1088/0004-637X/796/2/109}

\bibitem[{Aartsen {et~al.}(2014{\natexlab{b}})}]{Aartsen:2013vja}
---. 2014{\natexlab{b}}, JINST, 9, P03009,
  \dodoi{10.1088/1748-0221/9/03/P03009}

\bibitem[{Aartsen {et~al.}(2017{\natexlab{a}})}]{IceCube:2017trr}
---. 2017{\natexlab{a}}, Astrophys. J., 849, 67,
  \dodoi{10.3847/1538-4357/aa8dfb}

\bibitem[{Aartsen {et~al.}(2017{\natexlab{b}})}]{icecube_7year}
---. 2017{\natexlab{b}}, Astrophys. J., 835, 151,
  \dodoi{10.3847/1538-4357/835/2/151}

\bibitem[{Aartsen {et~al.}(2019)}]{IceCube:2019lzm}
---. 2019, Astrophys. J., 886, 12, \dodoi{10.3847/1538-4357/ab4ae2}

\bibitem[{Aartsen {et~al.}(2020{\natexlab{a}})}]{ic10yrs}
---. 2020{\natexlab{a}}, Phys. Rev. Lett., 124, 051103,
  \dodoi{10.1103/PhysRevLett.124.051103}

\bibitem[{Aartsen {et~al.}(2020{\natexlab{b}})}]{IceCube:2020svz}
---. 2020{\natexlab{b}}, Astrophys. J., 898, 117,
  \dodoi{10.3847/1538-4357/ab9fa0}

\bibitem[{Aartsen {et~al.}(2021)}]{IceCube-Gen2:2020qha}
---. 2021, J. Phys. G, 48, 060501, \dodoi{10.1088/1361-6471/abbd48}

\bibitem[{Abbasi {et~al.}(2009)Abbasi, Ackermann, Adams, Ahlers, Ahrens,
  Andeen, Auffenberg, Bai, Baker, Barwick, \& et~al.}]{icecube_daq}
Abbasi, R., Ackermann, M., Adams, J., {et~al.} 2009, Nuclear Instruments and
  Methods in Physics Research Section A: Accelerators, Spectrometers, Detectors
  and Associated Equipment, 601, 294–316, \dodoi{10.1016/j.nima.2009.01.001}

\bibitem[{Abbasi {et~al.}(2010)Abbasi, Abdou, Abu-Zayyad, Adams, Aguilar,
  Ahlers, Andeen, Auffenberg, Bai, Baker, \& et~al.}]{icecube_detector}
Abbasi, R., Abdou, Y., Abu-Zayyad, T., {et~al.} 2010, Nuclear Instruments and
  Methods in Physics Research Section A: Accelerators, Spectrometers, Detectors
  and Associated Equipment, 618, 139–152, \dodoi{10.1016/j.nima.2010.03.102}

\bibitem[{Abbasi {et~al.}(2022{\natexlab{a}})}]{IceCube:2022der}
Abbasi, R., {et~al.} 2022{\natexlab{a}}, Science, 378, 538,
  \dodoi{10.1126/science.abg3395}

\bibitem[{Abbasi {et~al.}(2022{\natexlab{b}})}]{IceCube:2021waz}
---. 2022{\natexlab{b}}, Astrophys. J., 926, 59,
  \dodoi{10.3847/1538-4357/ac3cb6}

\bibitem[{Abbasi {et~al.}(2022{\natexlab{c}})}]{IceCube:2022ccm}
---. 2022{\natexlab{c}}, Astrophys. J. Lett., 938, L11,
  \dodoi{10.3847/2041-8213/ac966b}

\bibitem[{Abbasi {et~al.}(2023{\natexlab{a}})}]{2023icgpdiffuse}
---. 2023{\natexlab{a}}, Science, 380, 1338, \dodoi{10.1126/science.adc9818}

\bibitem[{Abbasi {et~al.}(2023{\natexlab{b}})}]{IceCube:2022heu}
---. 2023{\natexlab{b}}, Astrophys. J. Lett., 945, L8,
  \dodoi{10.3847/2041-8213/acb933}

\bibitem[{Abdalla {et~al.}(2018)}]{HESS:2018pbp}
Abdalla, H., {et~al.} 2018, Astron. Astrophys., 612, A1,
  \dodoi{10.1051/0004-6361/201732098}

\bibitem[{{Abdalla} {et~al.}(2021){Abdalla}, {Aharonian}, {Ait Benkhali},
  {Ang{\"u}ner}, {Arcaro}, {Armand}, {Armstrong}, {Ashkar}, {Backes},
  {Baghmanyan}, {Barbosa Martins}, {Barnacka}, {Barnard}, {Becherini}, {Berge},
  {Bernl{\"o}hr}, {Bi}, {B{\"o}ttcher}, {Boisson}, {Bolmont}, {de Bonyde
  Lavergne}, {Breuhaus}, {Brose}, {Brun}, {Brun}, {Bryan}, {B{\"u}chele},
  {Bulik}, {Bylund}, {Caroff}, {Carosi}, {Chand}, {Chandra}, {Chen}, {Cotter},
  {Cury{\l}o}, {Damascene Mbarubucyeye}, {Davids}, {Davies}, {Deil}, {Devin},
  {Dirson}, {Djannati-Ata{\"\i}}, {Dmytriiev}, {Donath}, {Doroshenko},
  {Dreyer}, {Duffy}, {Dyks}, {Egberts}, {Eichhorn}, {Einecke}, {Emery},
  {Ernenwein}, {Feijen}, {Fegan}, {Fiasson}, {Fichet de Clairfontaine},
  {Fontaine}, {Funk}, {F{\"u}{\ss}ling}, {Gabici}, {Gallant}, {Giavitto},
  {Giunti}, {Glawion}, {Glicenstein}, {Gottschall}, {Grondin}, {Hahn}, {Haupt},
  {Hermann}, {Hinton}, {Hofmann}, {Hoischen}, {Holch}, {Holler}, {H{\"o}rbe},
  {Horns}, {Huber}, {Jamrozy}, {Jankowsky}, {Jankowsky}, {Jung-Richardt},
  {Kasai}, {Kastendieck}, {Katarzy{\'n}ski}, {Katz}, {Khangulyan},
  {Kh{\'e}lifi}, {Klepser}, {Klu{\'z}niak}, {Komin}, {Konno}, {Kosack},
  {Kostunin}, {Kreter}, {Lamanna}, {Lemi{\`e}re}, {Lemoine-Goumard}, {Lenain},
  {Leuschner}, {Levy}, {Lohse}, {Lypova}, {Mackey}, {Majumdar}, {Malyshev},
  {Malyshev}, {Marandon}, {Marchegiani}, {Marcowith}, {Mares},
  {Mart{\'\i}-Devesa}, {Marx}, {Maurin}, {Meintjes}, {Meyer}, {Mitchell},
  {Moderski}, {Mohrmann}, {Montanari}, {Moore}, {Morris}, {Moulin}, {Muller},
  {Murach}, {Nakashima}, {Nayerhoda}, {de Naurois}, {Ndiyavala}, {Niemiec},
  {Oakes}, {O'Brien}, {Odaka}, {Ohm}, {Olivera-Nieto}, {de Ona Wilhelmi},
  {Ostrowski}, {Panter}, {Panny}, {Parsons}, {Peron}, {Peyaud}, {Piel}, {Pita},
  {Poireau}, {Priyana Noel}, {Prokhorov}, {Prokoph}, {P{\"u}hlhofer}, {Punch},
  {Quirrenbach}, {Raab}, {Rauth}, {Reichherzer}, {Reimer}, {Reimer}, {Remy},
  {Renaud}, {Rieger}, {Rinchiuso}, {Romoli}, {Rowell}, {Rudak}, {Sahakian},
  {Sailer}, {Salzmann}, {Sanchez}, {Santangelo}, {Sasaki}, {Sch{\"a}fer},
  {Sch{\"u}ssler}, {Schutte}, {Schwanke}, {Seglar-Arroyo}, {Senniappan},
  {Seyffert}, {Shafi}, {Shapopi}, {Shiningayamwe}, {Simoni}, {Sinha}, {Sol},
  {Specovius}, {Spencer}, {Spir-Jacob}, {Stawarz}, {Sun}, {Steenkamp},
  {Stegmann}, {Steinmassl}, {Steppa}, {Takahashi}, {Tavernier}, {Taylor},
  {Terrier}, {Thiersen}, {Tiziani}, {Tluczykont}, {Tomankova}, {Trichard},
  {Tsirou}, {Tuffs}, {Uchiyama}, {van der Walt}, {van Eldik}, {van Rensburg},
  {van Soelen}, {Vasileiadis}, {Veh}, {Venter}, {Vincent}, {Vink}, {V{\"o}lk},
  {Wadiasingh}, {Wagner}, {Watson}, {Werner}, {White}, {Wierzcholska}, {Wong},
  {Yusafzai}, {Zacharias}, {Zanin}, {Zargaryan}, {Zdziarski}, {Zech}, {Zhu},
  {Zmija}, {Zorn}, {Zouari}, {{\.Z}ywucka}, {Albert}, {Alfaro}, {Alvarez},
  {Arteaga-Vel{\'e}zquez}, {Arunbabu}, {Avila Rojas}, {Belmont-Moreno},
  {BenZvi}, {Brisbois}, {Caballero-Mora}, {Capistr{\'a}n}, {Carrami{\~n}ana},
  {Casanova}, {Cotti}, {Cotzomi}, {Couti{\~n}o de Le{\'o}n}, {De la Fuente},
  {de Le{\'o}n}, {Diaz Hernandez}, {D{\'\i}az-V{\'e}lez}, {Dingus},
  {DuVernois}, {Durocher}, {Ellsworth}, {Engel}, {Espinoza}, {Fan},
  {Fern{\'a}ndez Alonso}, {Fraija}, {Galv{\'a}n-G{\'a}mez}, {Garcia},
  {Garc{\'\i}a-Gonz{\'a}lez}, {Garfias}, {Giacinti}, {Gonz{\'a}lez}, {Goodman},
  {Harding}, {Hernandez}, {Hona}, {Huang}, {Hueyotl-Zahuantitla},
  {H{\"u}ntemeyer}, {Iriarte}, {Jardin-Blicq}, {Joshi}, {Kieda}, {Lee},
  {Le{\'o}n Vargas}, {Linnemann}, {Longinotti}, {Luis-Raya}, {L{\'o}pez-Coto},
  {Malone}, {Martinez}, {Martinez-Castellanos}, {Mart{\'\i}nez-Castro},
  {Matthews}, {Miranda-Romagnoli}, {Morales-Soto}, {Moreno}, {Mostaf{\'a}},
  {Nayerhoda}, {Nellen}, {Newbold}, {Nisa}, {Noriega-Papaqui}, {Omodei},
  {Peisker}, {P{\'e}rez Araujo}, {P{\'e}rez-P{\'e}rez}, {Rho},
  {Rosa-Gonz{\'a}lez}, {Ruiz-Velasco}, {Salesa Greus}, {Sandoval}, {Schneider},
  {Schoorlemmer}, {Serna-Franco}, {Smith}, {Springer}, {Surajbali},
  {Tollefson}, {Torres}, {Torres-Escobedo}, {Turner}, {Ure{\~n}a-Mena},
  {Villase{\~n}or}, {Weisgarber}, {Willox}, {Zhou}, \& {HAWC
  Collaboration}}]{2021ApJ...917....6A}
{Abdalla}, H., {Aharonian}, F., {Ait Benkhali}, F., {et~al.} 2021, \apj, 917,
  6, \dodoi{10.3847/1538-4357/abf64b}

\bibitem[{Abeysekara {et~al.}(2020)}]{bib:astronomy_HAWC_GalacticPlane56TeV}
Abeysekara, A., {et~al.} 2020, Phys. Rev. Lett., 124, 021102,
  \dodoi{10.1103/PhysRevLett.124.021102}

\bibitem[{Abeysekara {et~al.}(2021)}]{bib:hawc_cygnus}
Abeysekara, A.~U., {et~al.} 2021, Nature Astron., 5, 465,
  \dodoi{10.1038/s41550-021-01318-y}

\bibitem[{Agostini {et~al.}(2020)}]{P-ONE:2020ljt}
Agostini, M., {et~al.} 2020, Nature Astron., 4, 913,
  \dodoi{10.1038/s41550-020-1182-4}

\bibitem[{Aharonian {et~al.}(2021)}]{bib:LHAASO_extended}
Aharonian, F., {et~al.} 2021, Phys. Rev. Lett., 126, 241103,
  \dodoi{10.1103/PhysRevLett.126.241103}

\bibitem[{Ahlers \& Murase(2014)}]{Ahlers:2013xia}
Ahlers, M., \& Murase, K. 2014, Phys. Rev. D, 90, 023010,
  \dodoi{10.1103/PhysRevD.90.023010}

\bibitem[{Aiello {et~al.}(2019)}]{KM3NeT:2018wnd}
Aiello, S., {et~al.} 2019, Astropart. Phys., 111, 100,
  \dodoi{10.1016/j.astropartphys.2019.04.002}

\bibitem[{Albert {et~al.}(2020{\natexlab{a}})}]{bib:3hwc_catalog}
Albert, A., {et~al.} 2020{\natexlab{a}}, Astrophys. J., 905, 76,
  \dodoi{10.3847/1538-4357/abc2d8}

\bibitem[{Albert {et~al.}(2020{\natexlab{b}})}]{bib:hawc_j2006}
---. 2020{\natexlab{b}}, Astrophys. J. Lett., 903, L14,
  \dodoi{10.3847/2041-8213/abbfae}

\bibitem[{Albert {et~al.}(2020{\natexlab{c}})}]{HAWC:2020nvc}
---. 2020{\natexlab{c}}, Astrophys. J. Lett., 896, L29,
  \dodoi{10.3847/2041-8213/ab96cc}

\bibitem[{Albert {et~al.}(2021{\natexlab{a}})}]{HAWC:2021dtl}
---. 2021{\natexlab{a}}, Astrophys. J. Lett., 911, L27,
  \dodoi{10.3847/2041-8213/abf4dc}

\bibitem[{Albert {et~al.}(2021{\natexlab{b}})}]{bib:hawc_j1825}
---. 2021{\natexlab{b}}, Astrophys. J. Lett., 907, L30,
  \dodoi{10.3847/2041-8213/abd77b}

\bibitem[{Amenomori {et~al.}(2021)}]{bib:tibet_cygnus}
Amenomori, M., {et~al.} 2021, Phys. Rev. Lett., 127, 031102,
  \dodoi{10.1103/PhysRevLett.127.031102}

\bibitem[{{Archer} {et~al.}(2016){Archer}, {Benbow}, {Bird}, {Buchovecky},
  {Buckley}, {Bugaev}, {Byrum}, {Cardenzana}, {Cerruti}, {Chen}, {Ciupik},
  {Collins-Hughes}, {Connolly}, {Eisch}, {Falcone}, {Feng}, {Finley},
  {Fleischhack}, {Flinders}, {Fortson}, {Furniss}, {Gillanders}, {Griffin},
  {Grube}, {Gyuk}, {H{\r{a}}kansson}, {Hanna}, {Holder}, {Humensky},
  {H{\"u}tten}, {Johnson}, {Kaaret}, {Kar}, {Kelley-Hoskins}, {Kertzman},
  {Kieda}, {Krause}, {Krennrich}, {Kumar}, {Lang}, {McArthur}, {McCann},
  {Meagher}, {Millis}, {Moriarty}, {Mukherjee}, {Nieto}, {Ong}, {Park},
  {Pelassa}, {Pohl}, {Popkow}, {Pueschel}, {Quinn}, {Ragan}, {Ratliff},
  {Reynolds}, {Richards}, {Roache}, {Rousselle}, {Santander}, {Sembroski},
  {Shahinyan}, {Smith}, {Staszak}, {Telezhinsky}, {Tucci}, {Tyler},
  {Vassiliev}, {Wakely}, {Weiner}, {Weinstein}, {Wilhelm}, {Williams},
  {Zitzer}, \& {Yusef-Zadeh}}]{2016ApJ...821..129A}
{Archer}, A., {Benbow}, W., {Bird}, R., {et~al.} 2016, \apj, 821, 129,
  \dodoi{10.3847/0004-637X/821/2/129}

\bibitem[{Blasi(2013)}]{Blasi_2013}
Blasi, P. 2013, Nuclear Physics B - Proceedings Supplements, 239-240, 140,
  \dodoi{10.1016/j.nuclphysbps.2013.05.023}

\bibitem[{{Bose} {et~al.}(2022){Bose}, {Chitnis}, {Majumdar}, \&
  {Shukla}}]{2022EPJST.231...27B}
{Bose}, D., {Chitnis}, V.~R., {Majumdar}, P., \& {Shukla}, A. 2022, European
  Physical Journal Special Topics, 231, 27,
  \dodoi{10.1140/epjs/s11734-022-00434-8}

\bibitem[{Braun {et~al.}(2008)Braun, Dumm, De~Palma, Finley, Karle, \&
  Montaruli}]{unbinned_llh}
Braun, J., Dumm, J., De~Palma, F., {et~al.} 2008, Astroparticle Physics, 29,
  299–305, \dodoi{10.1016/j.astropartphys.2008.02.007}

\bibitem[{Cao {et~al.}(2021)Cao, Aharonian, An, Bai, Bai, Bao, Bastieri, Bi,
  Bi, Cai, Cai, Cao, Chang, Chang, Chang, Chen, Chen, \& Chen}]{bib:lhaaso_12}
Cao, Z., Aharonian, F., An, Q., {et~al.} 2021, Nature, 594,
  \dodoi{10.1038/s41586-021-03498-z}

\bibitem[{Cao {et~al.}(2023)}]{LHAASO:2023gne}
Cao, Z., {et~al.} 2023, arXiv

\bibitem[{Denton {et~al.}(2017)Denton, Marfatia, \& Weiler}]{Denton:2017csz}
Denton, P.~B., Marfatia, D., \& Weiler, T.~J. 2017, JCAP, 08, 033,
  \dodoi{10.1088/1475-7516/2017/08/033}

\bibitem[{Fang {et~al.}(2022)Fang, Kerr, Blandford, Fleischhack, \&
  Charles}]{Fang:2022uge}
Fang, K., Kerr, M., Blandford, R., Fleischhack, H., \& Charles, E. 2022, Phys.
  Rev. Lett., 129, 071101, \dodoi{10.1103/PhysRevLett.129.071101}

\bibitem[{{Gabici} {et~al.}(2019){Gabici}, {Evoli}, {Gaggero}, {Lipari},
  {Mertsch}, {Orlando}, {Strong}, \& {Vittino}}]{2019IJMPD..2830022G}
{Gabici}, S., {Evoli}, C., {Gaggero}, D., {et~al.} 2019, International Journal
  of Modern Physics D, 28, 1930022, \dodoi{10.1142/S0218271819300222}

\bibitem[{G{\'o}rski {et~al.}(2005)G{\'o}rski, Hivon, Banday, Wandelt, Hansen,
  Reinecke, \& Bartelmann}]{2005ApJ...622..759G}
G{\'o}rski, K.~M., Hivon, E., Banday, A.~J., {et~al.} 2005, Astrophys. J., 622,
  759, \dodoi{10.1086/427976}

\bibitem[{Hooper \& Linden(2022)}]{PhysRevD.105.103013}
Hooper, D., \& Linden, T. 2022, Phys. Rev. D, 105, 103013,
  \dodoi{10.1103/PhysRevD.105.103013}

\bibitem[{Klein \& Nishina(1929)}]{Klein1929}
Klein, O., \& Nishina, Y. 1929, Zeitschrift f\"ur Physik, 52, 853,
  \dodoi{10.1007/BF01366453}

\bibitem[{{MAGIC Collaboration} {et~al.}(2020){MAGIC Collaboration}, {Acciari},
  {Ansoldi}, {Antonelli}, {Arbet Engels}, {Baack}, {Babi{\'c}}, {Banerjee},
  {Barres de Almeida}, {Barrio}, {Becerra Gonz{\'a}lez}, {Bednarek},
  {Bellizzi}, {Bernardini}, {Berti}, {Besenrieder}, {Bhattacharyya},
  {Bigongiari}, {Biland}, {Blanch}, {Bonnoli}, {Bo{\v{s}}njak}, {Busetto},
  {Carosi}, {Ceribella}, {Chai}, {Chilingaryan}, {Cikota}, {Colak}, {Colin},
  {Colombo}, {Contreras}, {Cortina}, {Covino}, {D'Elia}, {da Vela}, {Dazzi},
  {de Angelis}, {de Lotto}, {Delfino}, {Delgado}, {Depaoli}, {di Pierro}, {di
  Venere}, {Do Souto Espi{\~n}eira}, {Dominis Prester}, {Donini}, {Dorner},
  {Doro}, {Elsaesser}, {Fallah Ramazani}, {Fattorini}, {Fern{\'a}ndez-Barral},
  {Ferrara}, {Fidalgo}, {Foffano}, {Fonseca}, {Font}, {Fruck}, {Fukami},
  {Garc{\'\i}a L{\'o}pez}, {Garczarczyk}, {Gasparyan}, {Gaug}, {Giglietto},
  {Giordano}, {Godinovi{\'c}}, {Green}, {Guberman}, {Hadasch}, {Hahn},
  {Herrera}, {Hoang}, {Hrupec}, {H{\"u}tten}, {Inada}, {Inoue}, {Ishio},
  {Iwamura}, {Jouvin}, {Kerszberg}, {Kubo}, {Kushida}, {Lamastra}, {Lelas},
  {Leone}, {Lindfors}, {Lombardi}, {Longo}, {L{\'o}pez}, {L{\'o}pez-Coto},
  {L{\'o}pez-Oramas}, {Loporchio}, {Machado de Oliveira Fraga}, {Maggio},
  {Majumdar}, {Makariev}, {Mallamaci}, {Maneva}, {Manganaro}, {Mannheim},
  {Maraschi}, {Mariotti}, {Mart{\'\i}nez}, {Masuda}, {Mazin},
  {Mi{\'c}anovi{\'c}}, {Miceli}, {Minev}, {Miranda}, {Mirzoyan}, {Molina},
  {Moralejo}, {Morcuende}, {Moreno}, {Moretti}, {Munar-Adrover}, {Neustroev},
  {Nigro}, {Nilsson}, {Ninci}, {Nishijima}, {Noda}, {Nogu{\'e}s}, {N{\"o}the},
  {Nozaki}, {Paiano}, {Palacio}, {Palatiello}, {Paneque}, {Paoletti},
  {Paredes}, {Pe{\~n}il}, {Peresano}, {Persic}, {Prada Moroni}, {Prandini},
  {Puljak}, {Rhode}, {Rib{\'o}}, {Rico}, {Righi}, {Rugliancich}, {Saha},
  {Sahakyan}, {Saito}, {Sakurai}, {Satalecka}, {Schmidt}, {Schweizer},
  {Sitarek}, {{\v{S}}nidari{\'c}}, {Sobczynska}, {Somero}, {Stamerra}, {Strom},
  {Strzys}, {Suda}, {Suri{\'c}}, {Takahashi}, {Tavecchio}, {Temnikov},
  {Terzi{\'c}}, {Teshima}, {Torres-Alb{\`a}}, {Tosti}, {Tsujimoto}, {Vagelli},
  {van Scherpenberg}, {Vanzo}, {Vazquez Acosta}, {Vigorito}, {Vitale}, {Vovk},
  {Will}, \& {Zari{\'c}}}]{2020A&A...642A.190M}
{MAGIC Collaboration}, {Acciari}, V.~A., {Ansoldi}, S., {et~al.} 2020, \aap,
  642, A190, \dodoi{10.1051/0004-6361/201936896}

\bibitem[{Murase {et~al.}(2013)Murase, Ahlers, \& Lacki}]{Murase:2013rfa}
Murase, K., Ahlers, M., \& Lacki, B.~C. 2013, Phys. Rev. D, 88, 121301,
  \dodoi{10.1103/PhysRevD.88.121301}

\bibitem[{Sudoh \& Beacom(2023{\natexlab{a}})}]{Sudoh:2022sdk}
Sudoh, T., \& Beacom, J.~F. 2023{\natexlab{a}}, Phys. Rev. D, 107, 043002,
  \dodoi{10.1103/PhysRevD.107.043002}

\bibitem[{Sudoh \& Beacom(2023{\natexlab{b}})}]{Sudoh:2023qrz}
---. 2023{\natexlab{b}}, arXiv preprint.
\newblock \doarXiv{2305.07043}

\bibitem[{Sudoh {et~al.}(2021)Sudoh, Linden, \& Hooper}]{Sudoh:2021avj}
Sudoh, T., Linden, T., \& Hooper, D. 2021, JCAP, 08, 010,
  \dodoi{10.1088/1475-7516/2021/08/010}

\bibitem[{Suvorova {et~al.}(2021)}]{Baikal-GVD:2021ypx}
Suvorova, O.~V., {et~al.} 2021, PoS, ICRC2021, 946, \dodoi{10.22323/1.395.0946}

\bibitem[{{Tibet AS{$\gamma$} Collaboration} {et~al.}(2021){Tibet AS{$\gamma$}
  Collaboration}, Amenomori, Bao, Bi, Chen, Chen, Chen, Chen, Chen,
  {Cirennima}, Cui, {Danzengluobu}, Ding, Fang, Fang, Feng, Feng, Feng, Gao,
  Gou, Guo, Guo, He, He, Hibino, Hotta, Hu, Hu, Huang, Jia, Jiang, Jin,
  Kasahara, Katayose, Kato, Kato, Kawata, Kihara, Ko, Kozai, {Labaciren}, Le,
  Li, Li, Li, Lin, Liu, Liu, Liu, Liu, Liu, Lou, Lu, Meng, Munakata, Nakada,
  Nakamura, Nanjo, Nishizawa, Ohnishi, Ohura, Ozawa, Qian, Qu, Saito, Sakata,
  Sako, Shao, Shibata, Shiomi, Sugimoto, Takano, Takita, Tan, Tateyama, Torii,
  Tsuchiya, Udo, Wang, Wu, Xue, Yamamoto, Yang, Yokoe, Yuan, Zhai, Zhang,
  Zhang, Zhang, Zhang, Zhang, Zhang, Zhang, Zhao, {Zhaxisangzhu}, \&
  Zhou}]{tibetasgcollaborationFirstDetectionSubPeV2021}
{Tibet AS{$\gamma$} Collaboration}, Amenomori, M., Bao, Y.~W., {et~al.} 2021,
  Phys. Rev. Lett., 126, 141101, \dodoi{10.1103/PhysRevLett.126.141101}

\bibitem[{{Wakely} \& {Horan}(2008)}]{2008ICRC....3.1341W}
{Wakely}, S.~P., \& {Horan}, D. 2008, in International Cosmic Ray Conference,
  Vol.~3, International Cosmic Ray Conference, 1341--1344

\bibitem[{{Ward} \& {VERITAS Collaboration}(2010)}]{2010int..workE...5W}
{Ward}, J.~E., \& {VERITAS Collaboration}. 2010, in Eighth Integral Workshop.
  The Restless Gamma-ray Universe (INTEGRAL 2010), 5,
  \dodoi{10.22323/1.115.0005}

\end{thebibliography}
\bibliographystyle{aasjournal}
\end{document}